\documentclass[aps,prb,reprint,superscriptaddress,longbibliography,floatfix]{revtex4-2}

\usepackage[T1]{fontenc}
\usepackage[utf8]{inputenc}
\usepackage{lmodern}
\usepackage{microtype}
\usepackage{amsmath,amssymb,mathtools,bm}
\usepackage{graphicx}
\usepackage{xcolor}
\usepackage{hyperref}
\hypersetup{
  colorlinks=true,
  linkcolor=blue!50!black,
  citecolor=blue!50!black,
  urlcolor=blue!50!black,
  pdftitle={Phase-accumulating hyperfine strain sensing with rare-earth-ion phase memories},
  pdfauthor={Mustafa Gundogan},
  pdfkeywords={rare-earth ions, phase memory, Raman heterodyne, strain sensing, dynamical decoupling}
}

\usepackage{comment}
\newcommand{\dd}{\mathrm{d}}
\newcommand{\ii}{\mathrm{i}}
\newcommand{\ee}{\mathrm{e}}
\newcommand{\eps}{\epsilon}

\newcommand{\ket}[1]{|#1\rangle}
\newcommand{\bra}[1]{\langle #1|}

\newcommand{\Gs}{G_s}
\newcommand{\Nsb}{N_{\mathrm{sb}}}
\newcommand{\Neff}{N_{\mathrm{eff}}}
\definecolor{revisionred}{RGB}{175,28,38}
\ifdefined\markedversion
  \colorlet{revisioncolor}{revisionred}
\else
  \colorlet{revisioncolor}{black}
\fi
\DeclareRobustCommand{\rev}[1]{{\color{revisioncolor}#1}}
\newcommand{\revstart}{\color{revisioncolor}}
\newcommand{\revstop}{\color{black}}

\begin{document}

\title{Phase-accumulating hyperfine strain sensing with rare-earth-ion phase memories}
\author{Mustafa G\"undo\u{g}an}
\email{mustafa.guendogan@physik.hu-berlin.de}
\affiliation{Institut f\"ur Physik and Center for the Science of Materials Berlin (CSMB), Humboldt-Universit\"at zu Berlin, Newtonstr. 15, 12489 Berlin, Germany}
\date{September 25, 2026}

\begin{abstract}
\revstart
Long-lived hyperfine coherences in rare-earth-ion crystals provide controlled phase evolution windows for metrology. We use this resource to propose an experimental protocol for accessing strain shifts of ground state hyperfine transitions in non-Kramers rare-earth-ion doped (REID) materials, a quantity that is difficult to access directly with conventional spectroscopic methods. Starting from the crystal field Hamiltonian, we relate the response to strain-dependent effective quadrupole and Zeeman tensors, including changes in electronic wave functions, virtual electronic admixtures, and the bare nuclear quadrupole interaction. The protocol optically prepares a selected hyperfine class, uses a phase-controlled rf $\pi/2$ pulse to create the sensing coherence, and applies synchronized dynamical decoupling pulses during the phase evolution interval so that a coherent ac strain drive accumulates phase rather than averaging away. The accumulated phase is then retrieved by Raman heterodyne readout. The resulting framework provides a route to understanding strain-induced hyperfine couplings in non-Kramers REID systems.\revstop
\end{abstract}

\maketitle

\section{Introduction}
\revstart
Rare-earth-ion-doped (REID) crystals combine several properties that are unusual in a solid-state system. The optically active $4f$ electrons are shielded by filled outer electronic shells, giving rise to narrow optical transitions and long-lived hyperfine coherences, while the large inhomogeneous broadening of the optical lines allows spectral-class selection and optical pumping \cite{Macfarlane2002,Goldner2015}. These properties have made REID crystals important platforms for optical quantum memories and coherent light--matter interfaces. Optical excitations can be transferred to long-lived ground state coherences, and those spin-wave excitations can be protected by rf dynamical decoupling (DD) while preserving optical phase \cite{Lovric2013,Gundogan2015}. Depending on the material and operating point, coherence or storage times ranging from milliseconds to minutes or hours have been demonstrated \cite{Fraval2004,Heinze2013,Zhong2015,Ma2021,Ortu2022}.

The same separation of optical and hyperfine energy scales that makes these systems attractive for memories also creates an interesting spectroscopy problem. Strain changes the local crystal field, producing optical line shifts and broadening that can be measured by piezospectroscopy \cite{Galland2020,Zhang2020}. The corresponding response of a ground state hyperfine transition is much less direct. Optical line shifts do not by themselves determine this response. For non-Kramers ions such as Pr$^{3+}$ and Eu$^{3+}$, the hyperfine splitting is described by an effective spin Hamiltonian. Strain changes its parameters through the local electric field gradient and through changes in electronic eigenstates and virtual admixtures; the relative contributions depend on the ion and host. The hyperfine response can be much smaller than the optical response and can be difficult to isolate in a frequency domain measurement.

This difficulty is particularly clear when comparing the relevant energy scales. Hyperfine transitions lie in the MHz range, while the strain-induced changes of interest may be at the Hertz level. Optical transitions, by contrast, can undergo much larger strain-induced shifts and broadening. In an optical hole-burning measurement, one must then track small changes in the relative positions of hyperfine-related features despite potentially larger optical shifts and broadening; a common optical shift alone does not preclude such a differential measurement. An alternative is not to resolve the small frequency shift directly, but to let it act for a sufficiently long time and measure the phase that it accumulates. This principle already has a close experimental precedent: Macfarlane \emph{et al.} used Raman heterodyne spin echoes to resolve sub-Hertz electric field-induced shifts of rare-earth hyperfine transitions by converting the frequency change into an accumulated spin phase \cite{Macfarlane2014}.

This is closely related to the older idea of a \textit{phase memory}. Spin echoes were described in terms of phase memory well before the modern language of quantum information \cite{Mims1968}, and photon echoes in inhomogeneously broadened solids were later developed as frequency-selective optical data memories \cite{Mossberg1982}. In both cases, a controlled field writes phase information into a material coherence, the coherence retains and evolves that phase, and a later field reads it. Quantum optical memories impose the stronger requirement that an unknown optical state be stored and retrieved faithfully, but the same long-lived ensemble coherence can also be used more simply as a classical phase memory resource for metrology.

Here we use that resource to measure strain-induced shifts of ground state hyperfine transitions. A selected spectral class is first prepared by optical pumping, and a phase-controlled rf $\pi/2$ pulse writes a transverse coherence between two hyperfine states. During the interrogation interval, an oscillatory strain field modulates the hyperfine transition frequency. For an ac signal, the rf $\pi$ pulses are synchronized to the strain zero crossings, reversing the sign of the phase response when the strain changes sign. Successive half-cycles then contribute with the same sign and the strain-induced phase accumulates instead of averaging away. The same DD sequence suppresses slow frequency noise and defines the narrowband response of the sensor.

The accumulated spin phase is read optically by Raman heterodyne detection. In a three-level system, an optical read field together with the ground state coherence generates a Raman sideband; interference between this sideband and the optical carrier converts the spin phase into an optical beat phase \cite{Mlynek1983,Wong1983,Mitsunaga1985}. Raman heterodyne spectroscopy has been used for nuclear and electron-spin resonance, multipulse spin echoes, electric field shifts of rare-earth nuclear coherences, and microwave-to-optical conversion \cite{Holliday1990,Erickson1990,Macfarlane2014,FernandezGonzalvo2019,King2021}. We therefore treat the sensor as an rf-written, optically read phase memory.

\begin{figure*}[t]
\centering
\includegraphics[width=1\textwidth]{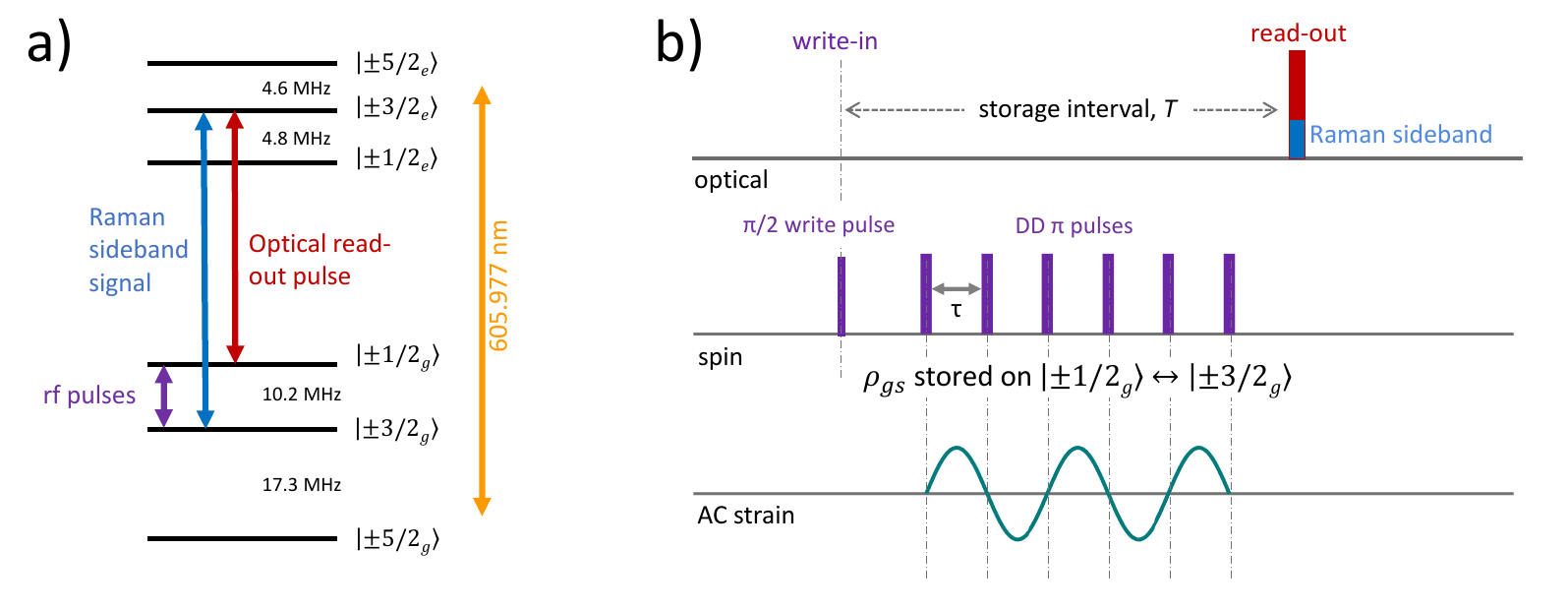}
\caption{\rev{Proposed strain measurement scheme.
(a) Relevant Pr$^{3+}$:Y$_2$SiO$_5$ hyperfine levels and optical transitions near 605.977 nm. The ground state transition $\ket{\pm1/2_g}\leftrightarrow\ket{\pm3/2_g}$ is controlled by rf fields. An optical read field on $\ket{\pm1/2_g}\leftrightarrow\ket{\pm3/2_e}$ generates a coherent Raman sideband on $\ket{\pm3/2_g}\leftrightarrow\ket{\pm3/2_e}$. (b) Timing sequence. Before the sequence shown, optical pumping polarizes the addressed spectral class, ideally into $\ket{\pm1/2_g}$. A phase-controlled rf $\pi/2$ pulse with phase $\phi_w$ creates the ground state coherence $\rho_{gs}$. During the storage interval, synchronized rf $\pi$ pulses provide dynamical decoupling while the ac strain contributes the accumulated phase $\Phi_{\eps}$. The final optical read field generates the Raman sideband, whose beat phase yields the Raman heterodyne phase $\phi_{\mathrm{RH}}$.}}
\label{fig:protocol}
\end{figure*}

The manuscript addresses four questions: which microscopic tensors determine the hyperfine strain response; how the strain-induced phase accumulates under synchronized DD; how large a transverse coherence can be prepared; and how precisely that phase can be read out. We then consider the anisotropic conversion between applied stress and local strain in low-symmetry crystals, electric-field and thermal systematics of piezo actuation, site symmetry and ZEFOZ operation, and the reconstruction of effective strain tensors from phase slopes measured on different hyperfine transitions.
\revstop

\section{Hyperfine phase memory protocol}
\revstart
\label{sec:protocol}

We consider a generic three-level implementation. Two long-lived hyperfine levels $\ket{g}$ and $\ket{s}$ form the sensing transition, while an optically excited state $\ket{e}$ provides optical pumping and readout. A complete experimental cycle consists of four operations.

First, narrowband optical pumping prepares a selected spectral class with populations $p_g$ and $p_s$ and population difference
\begin{equation}
P \equiv p_g-p_s.
\label{eq:P}
\end{equation}
Second, a resonant rf pulse of area $\theta=\pi/2$ and phase $\phi_w$ rotates the population difference into a transverse coherence and thereby writes the phase into the material. Third, the coherence evolves for a time $T$ while strain shifts the hyperfine transition. Refocusing or DD pulses produce a sign-changing sensitivity function $y(t)=\pm1$, suppress slow magnetic and frequency noise, and select the desired strain frequency. Finally, an optical read field on one leg of the $\Lambda$ system converts the ground state coherence into a Raman sideband on the other leg. The sideband interferes with the transmitted read carrier, or with a phase-related local oscillator, and the demodulated beat gives $\phi_{\mathrm{RH}}$.

For an even number of refocusing pulses, or after correcting their known parity, the retrieved phase in the convention specified in Appendix~\ref{app:writephase} has the form
\begin{equation}
\phi_{\mathrm{RH}}=\phi_w+\Phi_{\eps}+\Phi_c+\phi_0,
\label{eq:phase_transfer}
\end{equation}
where $\Phi_{\eps}$ is the strain phase, $\Phi_c$ contains known phases imposed by rf control, and $\phi_0$ is a fixed optical and electronic offset. A reference cycle with the actuator disabled, or a differential measurement between opposite strain phases, removes $\Phi_c+\phi_0$. Scanning $\phi_w$ must then produce a line of unit slope, while strain translates that line vertically by $\Phi_{\eps}$. This write-retain-read test is an operational verification that the device functions as a phase memory independently of an absolute Raman sideband calibration. Figure~\ref{fig:protocol} shows the level structure for Pr$^{3+}$:Y$_2$SiO$_5$ and relevant experimental sequence.
\revstop

\textcolor{black}{Another option is to implement a full, optical input-output type memory operation~\cite{Lovric2013, Ortu2022, Moldes2026} with DD at zero B-field. A comparison with such an optical memory operation would give valuable insights because the two approaches make almost opposite tradeoffs. For an optical input--output memory, the relevant comparison is the usual linear-storage regime, in which the input field is kept weak enough not to modify the prepared medium. In an atomic frequency comb (AFC) memory, a strong input would redistribute the prepared population and distort or wash out the comb, while in electromagnetically induced transparency (EIT)- or Raman-type memories it would invalidate the weak-probe approximation and lead to nonlinear storage dynamics. We therefore restrict the comparison to weak optical inputs, for which the stored spin wave corresponds to a small Bloch-sphere tilt~\cite{Sevincli2026}. An rf $\pi/2$ pulse, by contrast, can rotate the already prepared hyperfine population difference close to the equator without requiring a strong optical storage field. The advantage appears at readout: an optimized memory can map a large fraction of the stored excitation back into a well-defined optical output mode, and high storage-and-retrieval efficiencies have been demonstrated in rare-earth crystals \cite{Hedges2010}. In the direct rf-written scheme, by contrast, a $\pi/2$ pulse can rotate the prepared population difference to the equator and therefore create the largest mean hyperfine coherence available from that class. The optical readout is then not a direct retrieval of the stored excitation but a driven Raman conversion of the existing spin coherence into a sideband. In the weak-read limit the sideband field is linear in the read field and in $\rho_{gs}$, but increasing the read intensity eventually brings saturation, power broadening, optical pumping, reabsorption, and backaction. Thus the rf--Raman scheme gains coherence amplitude at the preparation step, whereas an optical input--output memory has the cleaner and potentially much more efficient optical retrieval (see Appendix~\ref{app:write} for a quantitative comparison).}

\section{Microscopic origin of the hyperfine strain coupling}
\label{sec:micro}

For a chosen transition, we describe the strain response by a frequency slope $\Gs$. This section identifies the microscopic object represented by that slope and distinguishes it from an optical piezospectroscopic coefficient.

\subsection{Crystal field strain Hamiltonian}
\revstart

Within a fixed electronic angular momentum multiplet, as used for Pr$^{3+}$, a convenient microscopic Hamiltonian is
\begin{equation}
\begin{aligned}
H={}&H_{\mathrm{FI}}+H_{\mathrm{CF}}+A_J\bm I\cdot\bm J+H_Q^{(n)}+H_{\eps}\\
&+g_J\mu_B\bm B\cdot\bm J-g_n\mu_N\bm B\cdot\bm I.
\end{aligned}
\label{eq:Hmicro}
\end{equation}
The crystal field is conventionally expanded in Stevens operators \cite{Stevens1952,Hutchings1964},
\begin{equation}
H_{\mathrm{CF}}=\sum_{kq}B_q^k O_q^k.
\end{equation}
Using $\eps_a=(\eps_{xx},\eps_{yy},\eps_{zz},\eps_{yz},\eps_{xz},\eps_{xy})$, without engineering-shear factors of two, the linear strain perturbation may be written
\begin{equation}
H_{\eps}=\sum_a \eps_a V_a,
\qquad
V_a=\sum_{kq}\gamma_a^{kq}O_q^k,
\label{eq:Hstrain}
\end{equation}
where $\gamma_a^{kq}=\partial B_q^k/\partial\eps_a$ are ion- and host-specific crystal field strain parameters.

For the Pr$^{3+}$ and Eu$^{3+}$ systems considered here, the relevant electronic crystal field states are singlets. Projecting the electronic degrees of freedom to a selected singlet gives the effective hyperfine Hamiltonian in frequency units \cite{Teplov1968,Longdell2002,Longdell2006,GuillotNoel2009}
\begin{equation}
\frac{H_s}{h}=B_iM_{ij}(\eps)I_j+I_iQ_{ij}(\eps)I_j,
\label{eq:Heff}
\end{equation}
with repeated Cartesian indices summed. Standard second-order expressions for the enhanced nuclear Zeeman tensor $M$, the effective quadrupole tensor $Q$, and the electronic susceptibility tensor $\Lambda$ are collected in Appendix~\ref{app:projection}. Strain changes both the crystal field eigenvectors and the electronic energy denominators entering these tensors. Consequently it changes $M$ and $Q$. Direct strain dependence of the bare nuclear quadrupole interaction or of hyperfine constants can be included in the corresponding effective derivatives and does not alter the phase protocol. For Eu$^{3+}$, whose ground state has $J=0$, inter-multiplet electronic matrix elements must replace the fixed-$J$ notation in the microscopic reduction; the effective form in Eq.~\eqref{eq:Heff} remains applicable (Appendix~\ref{app:projection}). Neither case is determined by an optical line shift coefficient alone.

\revstop

\subsection{Effective transition slope}

Expanding the tensors about an operating point gives
\begin{align}
Q_{ij}(\eps)&=Q_{ij}^{(0)}+\sum_aK_{ij}^{(Q,a)}\eps_a+O(\eps^2),\\
M_{ij}(\eps)&=M_{ij}^{(0)}+\sum_aK_{ij}^{(M,a)}\eps_a+O(\eps^2).
\end{align}
Here $K_{ij}^{(Q,a)}=\partial Q_{ij}/\partial\eps_a$ and $K_{ij}^{(M,a)}=\partial M_{ij}/\partial\eps_a$, evaluated at the operating point. Physically, these tensors describe how a given strain component changes the effective quadrupole and Zeeman couplings of the hyperfine manifold. At zero magnetic field, define the sensing transition frequency as $\nu_s=(E_s-E_g)/h$, where $\ket{g}$ and $\ket{s}$ are eigenstates of $\bm I\cdot Q^{(0)}\cdot\bm I$. First-order perturbation theory gives
\begin{equation}
\begin{aligned}
G_{s,a}(0)&=\frac{\partial\nu_s}{\partial\eps_a}\\
&=\bra{s}I_iK_{ij}^{(Q,a)}I_j\ket{s}
 -\bra{g}I_iK_{ij}^{(Q,a)}I_j\ket{g}.
\end{aligned}
\label{eq:Gsa}
\end{equation}
For a calibrated one-parameter strain pattern $\eps_a=\xi_a\eps$, the measured scalar coefficient is
\begin{equation}
\Gs(0)=\sum_a\xi_aG_{s,a}(0).
\label{eq:Gs}
\end{equation}
At finite field there is an additional contribution from $K^{(M,a)}$. The scalar $\Gs$ is therefore a projection of an effective hyperfine strain tensor. A single measurement does not determine one microscopic $\gamma_a^{kq}$; recovering those coefficients requires several transitions, strain geometries, and a crystal field model.

The expressions above identify the strain coefficient to be measured, but they do not fix its magnitude without a microscopic strain model. For the estimates below, we therefore use a simple scaling argument. If one virtual electronic energy denominator $\Delta_{\rm el}$ dominates the pseudoquadrupole contribution, the corresponding hyperfine splitting scales approximately as
\begin{equation}
\nu_s\sim\frac{A_J^2}{h\Delta_{\rm el}}.
\end{equation}
A strain-induced change of the electronic level spacing then gives
\begin{equation}
G_s=\frac{\partial\nu_s}{\partial\eps}
\sim
\nu_s\frac{1}{\Delta_{\rm el}}
\frac{\partial\Delta_{\rm el}}{\partial\eps}.
\label{eq:strain_scaling}
\end{equation}
Rare-earth optical strain and stress measurements give electronic frequency shift scales of order several to a few tens of THz/strain; we take $h^{-1}|\partial\Delta_{\rm el}/\partial\eps|\sim6$--$30$ THz/strain as a representative range \cite{Galland2020,Zhang2020}. Together with $\nu_s\sim10$--$100$ MHz and $\Delta_{\rm el}/h\sim1$--$10$ THz, this gives the working estimate $|G_s|\sim10^7$--$10^9$ Hz/strain. This estimate refers only to the virtual-admixture contribution; direct strain dependence of the bare quadrupole interaction and strain-induced changes of the electronic matrix elements can modify both the magnitude and sign of $G_s$. For scale, a strain amplitude $\eps=10^{-8}$ together with $|G_s|=10^8$ Hz/strain corresponds to a hyperfine shift of only $|\delta\nu_s|=|G_s\eps|=1$ Hz.
\subsection{Experimental benchmark from hyperfine Stark spectroscopy}
\revstart
\label{sec:benchmark}

Macfarlane \emph{et al.} provide a direct precedent for resolving small hyperfine shifts through phase accumulation \cite{Macfarlane2014}. Their Eu:YSO experiment used Stark-modulated quadrupole echoes with Raman heterodyne detection. The measured coefficients were $0.42$ and $1.0$ Hz cm/V for the 34.54- and 46.20-MHz transitions, compared with an optical Stark coefficient of approximately $27$ kHz cm/V.

These numbers also permit an order-of-magnitude comparison with a fractional-electronic-energy-scale estimate. If a single virtual electronic denominator dominates, this estimate reads
\begin{equation}
k_E^{(\mathrm{hf})}\sim \nu_s\frac{k_E^{(\mathrm{opt})}}{\Delta_{\mathrm{el}}/h}.
\label{eq:stark_scaling}
\end{equation}
Using $\nu_s=10$--$100$ MHz and $\Delta_{\mathrm{el}}/h=1$--$10$ THz gives $k_E^{(\mathrm{hf})}\sim0.03$--$3$ Hz cm/V, bracketing the measured values. This agreement is a scale comparison, not a validation of that mechanism: Macfarlane \emph{et al.} attribute the Eu:YSO nuclear Stark response mainly to the change in electric field gradient acting on the bare quadrupole moment, with only a small pseudoquadrupole contribution. Nor does the comparison establish a strain coefficient.

The experiment resolved a $0.6$-Hz change through Stark-induced echo modulation, not by locating a spectral line with sub-Hertz precision. Its rf interpulse delay was $10$ ms and the echo traces were typically averaged over 150 shots at 5 Hz. As a demonstrated frequency resolution, rather than a single-shot or per-root-Hertz sensitivity, this corresponds conditionally to
\begin{equation}
\delta\eps_{\mathrm{lit}}\simeq\frac{0.6\ \mathrm{Hz}}{|\Gs|}
=6\times10^{-8}\text{--}6\times10^{-10}
\label{eq:lit_benchmark}
\end{equation}
for $|\Gs|=10^7$--$10^9$ Hz/strain. This is an experimental benchmark for phase-based frequency readout; the proposed strain measurement still requires an independent strain calibration.
\revstop

\subsection{Finite-field and ZEFOZ operation}
\revstart
At a finite operating field, the eigenstates and the transition slope must be evaluated from the complete effective Hamiltonian in Eq.~\eqref{eq:Heff}. A zero-first-order-Zeeman (ZEFOZ) point satisfies
\begin{equation}
\nabla_{\bm B}\nu_s(\bm B_Z,\eps=0)=0,
\end{equation}
but this condition does not imply
\begin{equation}
\left.\frac{\partial\nu_s}{\partial\eps_a}\right|_{\bm B_Z}=0.
\end{equation}
Magnetic field and strain differentiate different parameters of the effective Hamiltonian. ZEFOZ operation can therefore suppress first-order magnetic noise while retaining a finite strain coefficient $G_{s,a}(\bm B_Z)$.

Strain can also move the magnetic sweet spot. Let $H_B$ be the Hessian of the transition frequency with respect to magnetic field at $\bm B_Z$. To first order in strain, the displacement that restores the ZEFOZ condition is
\begin{equation}
\delta\bm B_Z=-H_B^{-1}\sum_a\nabla_{\bm B}G_{s,a}(\bm B_Z)\eps_a,
\label{eq:ZEFOZshift}
\end{equation}
provided $H_B$ is invertible in the relevant subspace. The phase slope at fixed field and the strain-induced displacement of the ZEFOZ point are distinct observables.
\revstop

\section{Phase accumulation and DD filter functions}
\label{sec:filter}

\revstart
For one selected hyperfine transition, the strain-dependent two-level Hamiltonian can be written
\begin{equation}
\begin{aligned}
\frac{H_{gs}(t)}{\hbar}
&=-\frac{1}{2}\bigl[\omega_{gs}^{(0)}+2\pi\Gs\eps(t)
 +\delta\omega_n(t)\bigr]\sigma_z\\
&\quad+\frac{H_c(t)}{\hbar},
\end{aligned}
\end{equation}
Here $\sigma_z=\ket{g}\bra{g}-\ket{s}\bra{s}$ and $\omega_{gs}^{(0)}=2\pi\nu_s$; $\delta\omega_n$ denotes unwanted frequency noise and $H_c$ contains the rf control.

\textcolor{black}{In the toggling frame of ideal instantaneous $\pi$ pulses, the strain contribution is
\begin{equation}
H_{\eps}^{\rm tog}(t)=-\hbar\pi\Gs\,y(t)\eps(t)\sigma_z,
\end{equation}
where $y(t)=\pm1$ changes sign after each refocusing pulse. Since $H_{\eps}^{\rm tog}(t)$ is proportional to $\sigma_z$ at all times,
\begin{equation}
[H_{\eps}^{\rm tog}(t),H_{\eps}^{\rm tog}(t')]=0.
\end{equation}
The time ordering is therefore trivial, and the ground-state coherence evolves as
\begin{equation}
\rho_{gs}(T)=\rho_{gs}(0)\exp\!\left[\ii\,2\pi\Gs\int_0^T y(t)\eps(t)\dd t\right].
\label{eq:rho_phase_accumulation}
\end{equation}
The accumulated strain phase is thus}
\begin{equation}
\Phi_{\eps}(T)=2\pi\Gs\int_0^T y(t)\eps(t)\dd t,
\label{eq:phasefunctional}
\end{equation}
\revstop
with $y(t)=\pm1$. For a Fourier component $\eps(t)=\operatorname{Re}[\eps_\omega\ee^{-\ii\omega t}]$, define
\begin{equation}
Y(\omega)=\int_0^Ty(t)\ee^{-\ii\omega t}\dd t.
\label{eq:Y}
\end{equation}
Then
\begin{equation}
\Phi_{\eps}=2\pi\Gs\operatorname{Re}[\eps_\omega Y(\omega)].
\label{eq:phaseY}
\end{equation}
The same $Y(\omega)$ determines the response to coherent strain and, through $|Y(\omega)|^2$, the susceptibility to frequency noise \cite{Cywinski2008,Degen2017}.

For free evolution, $y(t)=1$ and
\begin{equation}
Y_0(\omega)=T\ee^{-\ii\omega T/2}\operatorname{sinc}\left(\frac{\omega T}{2}\right),
\end{equation}
where $\operatorname{sinc}x=\sin x/x$. A Hahn echo suppresses quasistatic shifts and responds near $\omega\sim\pi/T$. An equally spaced pulse train gives a narrow passband near half the $\pi$-pulse repetition rate. \textcolor{black}{The same multipulse lock-in principle is well established in ac spin sensing \cite{deLange2011}.} For a sinusoidal strain $\eps(t)=\eps_0\cos(\omega_mt+\phi_m)$ and an ideal square-wave sensitivity synchronized to the sign of the strain, the large-cycle limit gives
\begin{equation}
|\Phi_{\mathrm{DD,res}}|\simeq4|\Gs|\eps_0T.
\label{eq:matchedphase}
\end{equation}

\begin{figure*}[t]
\centering
\includegraphics[width=1\textwidth]{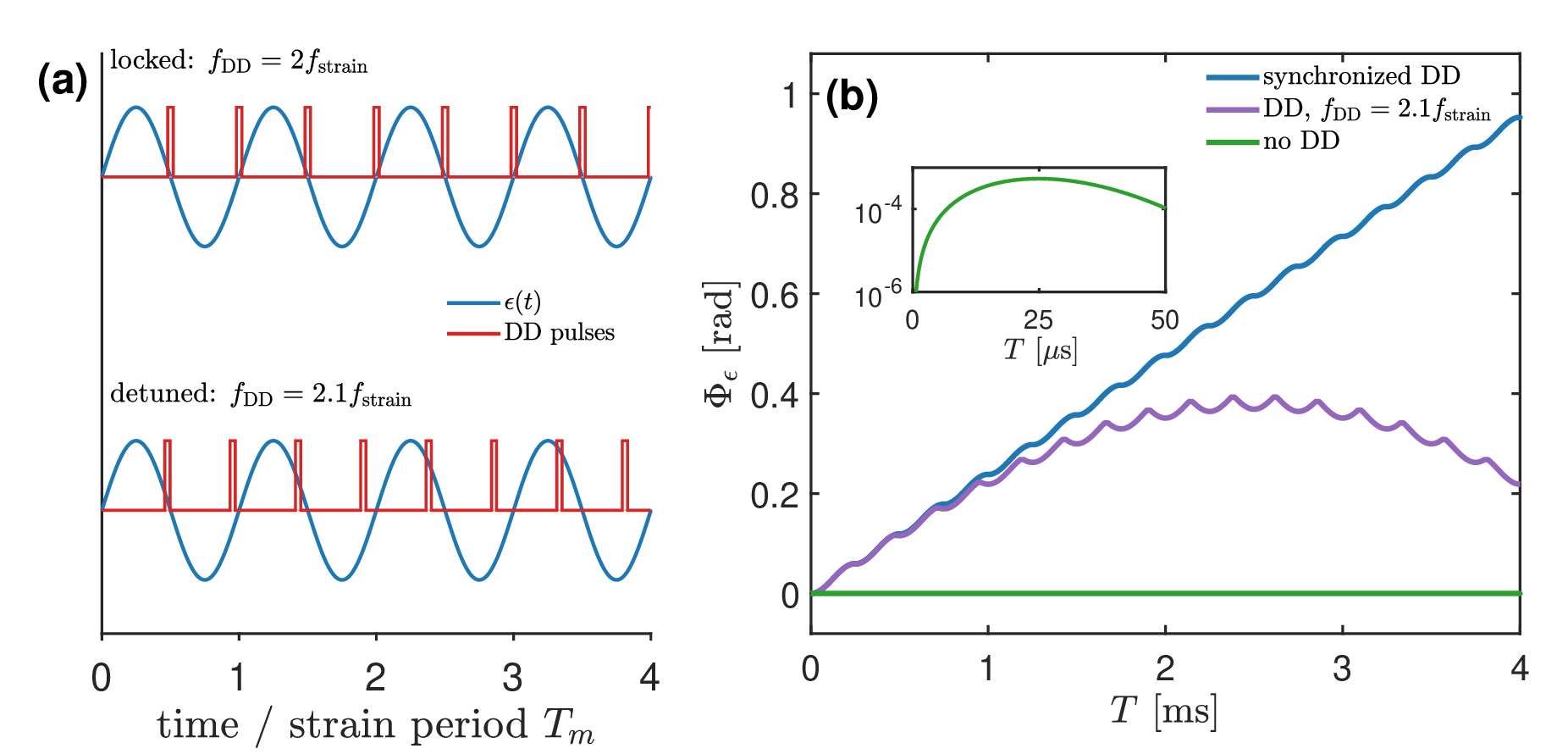}
\caption{\rev{Synchronization and phase accumulation. (a) The normalized ac strain (blue) and finite-width rf $\pi$-pulse envelope (red), drawn in the same waveform convention as the accompanying time-domain simulation. Here $f_{\rm DD}$ denotes the $\pi$-pulse repetition rate, so the locked condition is $f_{\rm DD}=2f_{\rm strain}$; the lower row shows the slip produced by the 5\% mismatch $f_{\rm DD}=2.1f_{\rm strain}$. The illustrated pulse duty cycle is $t_\pi f_{\rm DD}=0.08$. During each pulse the sensing function is set to zero as a pulse-blanking approximation, rather than a full driven Bloch-equation treatment. (b) Representative accumulated strain phase for $G_s\epsilon_0=60$ Hz and $f_{\rm strain}=2$ kHz using the same finite-pulse model. Synchronized DD produces nearly linear growth over the plotted $0$--$4$ ms interval, whereas the detuned sequence gives a beat and reduced net phase. The green trace and its $0$--$25\,\mu\mathrm{s}$ inset instead show the small-phase coherent signal $C_0(T)\Phi_{\eps,0}(T)$, with contrast $C_0(T)=\exp[-(T/25\,\mu\mathrm{s})^2]$, that shows the decay with $T_2^* \sim 25~\mu s$.}}
\label{fig:phase_accumulation}
\end{figure*}

\rev{For scale, $|\Gs\eps_0|=1$ Hz gives $|\Phi_{\eps}|\simeq0.04$ rad after $T=10$ ms in this ideal matched limit. Resolving that phase requires sufficient surviving coherence and readout precision. Figure~\ref{fig:phase_accumulation}(b) instead uses $|\Gs\eps_0|=60$ Hz: Eq.~\eqref{eq:matchedphase} gives $0.96$ rad at $T=4$ ms, reduced slightly to about $0.95$ rad by the finite-pulse model.}

For strain that is effectively constant over the interrogation time and no refocusing,
\begin{equation}
|\Phi_{\mathrm{free}}|\simeq2\pi|\Gs|\eps_0T.
\end{equation}
\textcolor{black}{DD therefore has two functions in this measurement: it preserves the coherence and sets the temporal weighting with which the strain contributes to the accumulated phase.}

\revstart
Real rf pulses introduce finite duration, detuning, amplitude error, and spatial Rabi-frequency inhomogeneity. We therefore write the surviving ensemble coherence as
\begin{equation}
\rho_{gs}(T)=\rho_{gs}(0)C(T,n_\pi)\ee^{\ii[\Phi_{\eps}(T)+\Phi_c(T)]},
\label{eq:contrast}
\end{equation}
where $C\leq1$ is the measured contrast after $n_\pi$ control pulses. Phase-robust sequences such as KDD or appropriately phased XY families are preferable when the initial transverse phase is deliberately scanned, because an imperfect CPMG train preferentially preserves one spin component \cite{Lovric2013,Souza2011}.
\revstop

\textcolor{black}{Figure~\ref{fig:phase_accumulation} shows why synchronization matters. When the sign changes of $y(t)$ coincide with the zero crossings of the strain, the product $y(t)\eps(t)$ keeps the same sign and successive half-cycles add constructively, giving an approximately linear phase growth. If the DD frequency is slightly detuned, the pulse train gradually slips in phase relative to the strain. The contributions then cease to add with the same sign, producing the beat-like response in Fig.~\ref{fig:phase_accumulation}(b); beyond the first maximum, later parts of the evolution can partially cancel phase accumulated earlier. A longer coherence time therefore does not by itself imply a larger signal unless the DD sequence remains phase matched to the strain. On the other hand, without DD, the sinusoidal strain is integrated with $y(t)=1$, so positive and negative half-cycles largely cancel and the accumulated phase remains bounded rather than growing with time. In addition, inhomogeneous broadening suppresses the observable ensemble coherence on the short $T_2^*$ scale. The $0$--$25\,\mu\mathrm{s}$ inset illustrates these two effects: the bare strain phase is small and oscillatory, while the plotted coherent signal $C_0(T)\Phi_{\eps,0}(T)$ rapidly disappears as the ensemble dephases.} \textcolor{black}{The corresponding ensemble dynamics on the Bloch sphere are shown in Supplemental Videos~1 and 2 for the synchronized and $5\%$-detuned DD sequences, respectively~\cite{Videos_SM}. The simulations use the same pulse timing as Fig.~\ref{fig:phase_accumulation} and show the individual pseudospins and collective Bloch vector during the strain-induced phase evolution.}

\section{Raman heterodyne phase readout}
\revstart
\label{sec:readout}

Assume that the optical read field drives $\ket{g}\leftrightarrow\ket{e}$ with complex Rabi frequency $\Omega_r$ and detuning $\Delta_r$. To first order in the ground state coherence, the optical coherence on the other leg is
\begin{equation}
\rho_{es}^{(1)}\simeq\frac{-\ii\Omega_r}{2(\gamma_{es}+\ii\Delta_{es})}\rho_{gs},
\label{eq:rhose}
\end{equation}
where $\gamma_{es}$ and $\Delta_{es}$ are the relevant optical decoherence rate and sideband detuning. The generated polarization radiates at $\omega_r\pm\omega_{gs}$. Propagation, optical depth, mode overlap, inhomogeneous broadening, and phase matching can be collected into a complex conversion coefficient $\mathcal K$,
\begin{equation}
E_{\mathrm{sb}}=\mathcal K E_r\Neff\rho_{gs}.
\label{eq:Esb}
\end{equation}
Equation~\eqref{eq:Esb} is a small-signal relation. For sensitivity estimates it is safer to use the experimentally measured coherent beat and define an \emph{equivalent detected sideband photon number} $\Nsb$: the photon number that a coherent sideband in the analyzed temporal and spectral mode would have if separated from the carrier. It can be inferred from the beat amplitude, the carrier photon number, and the measured visibility even when the sideband is not physically filtered from the carrier.

The sideband phase contains the optical read phase and the spin phase. Beating it against the transmitted read carrier cancels the common optical phase and gives
\begin{equation}
\phi_{\mathrm{RH}}=\arg(\rho_{gs})+\phi_{\mathcal K},
\end{equation}
where the fixed phase $\phi_{\mathcal K}$ includes detuning, propagation, and electronic demodulation. Optical readout also allows the magnetic field operating point or selected transition to be changed without requiring a high-$Q$ microwave detection resonator to be retuned to every transition.

For a carrier or local oscillator containing $N_{\mathrm{LO}}$ detected photons and a sideband containing $\Nsb$, the mean detected photon number versus relative phase is
\begin{equation}
N(\phi)=N_{\mathrm{LO}}+\Nsb+2\mathcal V\sqrt{N_{\mathrm{LO}}\Nsb}\cos\phi,
\end{equation}
where $0\leq\mathcal V\leq1$ accounts for mode overlap and coherence. In the strong-LO limit, the beat amplitude and LO shot noise both scale as $\sqrt{N_{\mathrm{LO}}}$; the LO therefore disappears from the quantum-limited phase precision. The optical contribution is
\begin{equation}
\delta\phi_{\mathrm{opt}}=\frac{\kappa_{\mathrm{det}}}{\mathcal V\sqrt{\Nsb}}.
\label{eq:dphiopt}
\end{equation}
The value $\kappa_{\mathrm{det}}=1/2$ is the ideal phase sensitive homodyne limit, requiring an optical local oscillator matched to the sideband frequency and measured quadrature. Conventional carrier--sideband Raman heterodyne gives $\kappa_{\mathrm{det}}=1/\sqrt{2}$, including the image-band penalty; electronic phase locking alone does not remove that penalty \cite{Anai2024}. \rev{Figure~\ref{fig:sensitivity} uses the ideal homodyne value.} For conventional Raman heterodyne, the optical noise-limited uncertainties increase by $\sqrt{2}$. Detection loss is included because $\Nsb$ is defined after detection.

The independent-spin projection floor of an rf-written coherent spin state is approximately
\begin{equation}
\delta\phi_{\mathrm{SPN}}\simeq\frac{1}{|P|C(T)\sqrt{\Neff}}.
\label{eq:SPN}
\end{equation}
Combining optical, spin-projection, and technical noise gives
\begin{equation}
\delta\phi^2\simeq
\frac{\kappa_{\mathrm{det}}^2}{\mathcal V^2\Nsb}
+\frac{1}{P^2C^2\Neff}
+\delta\phi_{\mathrm{tech}}^2.
\label{eq:totalphase}
\end{equation}
The first two terms are limiting contributions in an independent-spin, coherent-sideband model; a detailed microscopic readout model may correlate them. They become equal at
\begin{equation}
N_{\mathrm{sb}}^{\mathrm{cross}}
=\frac{\kappa_{\mathrm{det}}^2P^2C^2\Neff}{\mathcal V^2}.
\end{equation}
For ideal homodyne readout with $\mathcal V=1$ this is $P^2C^2\Neff/4$; for conventional Raman heterodyne it is $P^2C^2\Neff/2$. Reaching that point can require a large coherent sideband; in many experiments laser phase noise, vibration, rf phase drift, and actuator noise will become relevant first.
\revstop

\section{Strain sensitivity and benchmarks}
\revstart
\label{sec:sensitivity}

From Eqs.~\eqref{eq:phaseY} and \eqref{eq:totalphase}, the single-shot uncertainty of a strain quadrature aligned with the filter response is
\begin{equation}
\delta\eps(\omega)=\frac{\delta\phi}{2\pi|\Gs Y(\omega)|}.
\label{eq:depsgeneral}
\end{equation}
For a coherent ac strain field at the center of the DD filter passband, corresponding here to \(f_{\rm DD}=2f_{\rm strain}\), Eq.~\eqref{eq:matchedphase} gives
\begin{equation}
\delta\eps_{\mathrm{DD}}=\frac{\delta\phi}{4|\Gs|T}.
\label{eq:depsDD}
\end{equation}
If optical shot noise dominates,
\begin{equation}
\delta\eps_{\mathrm{DD,opt}}(T)
=\frac{\kappa_{\mathrm{det}}}
{4\mathcal V|\Gs|T C(T)\sqrt{N_{\mathrm{sb}}(0)}}.
\label{eq:depsopt}
\end{equation}
Here $N_{\mathrm{sb}}(0)$ is the equivalent detected sideband photon number extrapolated to zero storage time for the same preparation and read pulse. Equations~\eqref{eq:contrast} and \eqref{eq:Esb} give
\begin{equation}
N_{\mathrm{sb}}(T)=N_{\mathrm{sb}}(0)C^2(T).
\label{eq:Nsbdecay}
\end{equation}
Equation~\eqref{eq:depsopt} is therefore equivalent to the previous expression written using the sideband photon number actually detected at time $T$. For ideal phase sensitive homodyne readout it reduces to $1/[8\mathcal V|\Gs|T C(T)\sqrt{N_{\mathrm{sb}}(0)}]$. Thus, as in our recent analysis of memory-assisted squeezed light velocimetry~\cite{Gundogan2026_velo}, increasing the storage time improves the sensitivity only while the coherence \(C(T)\) remains sufficiently large.

\begin{figure}[!t]
\centering
\includegraphics[width=0.485\textwidth]{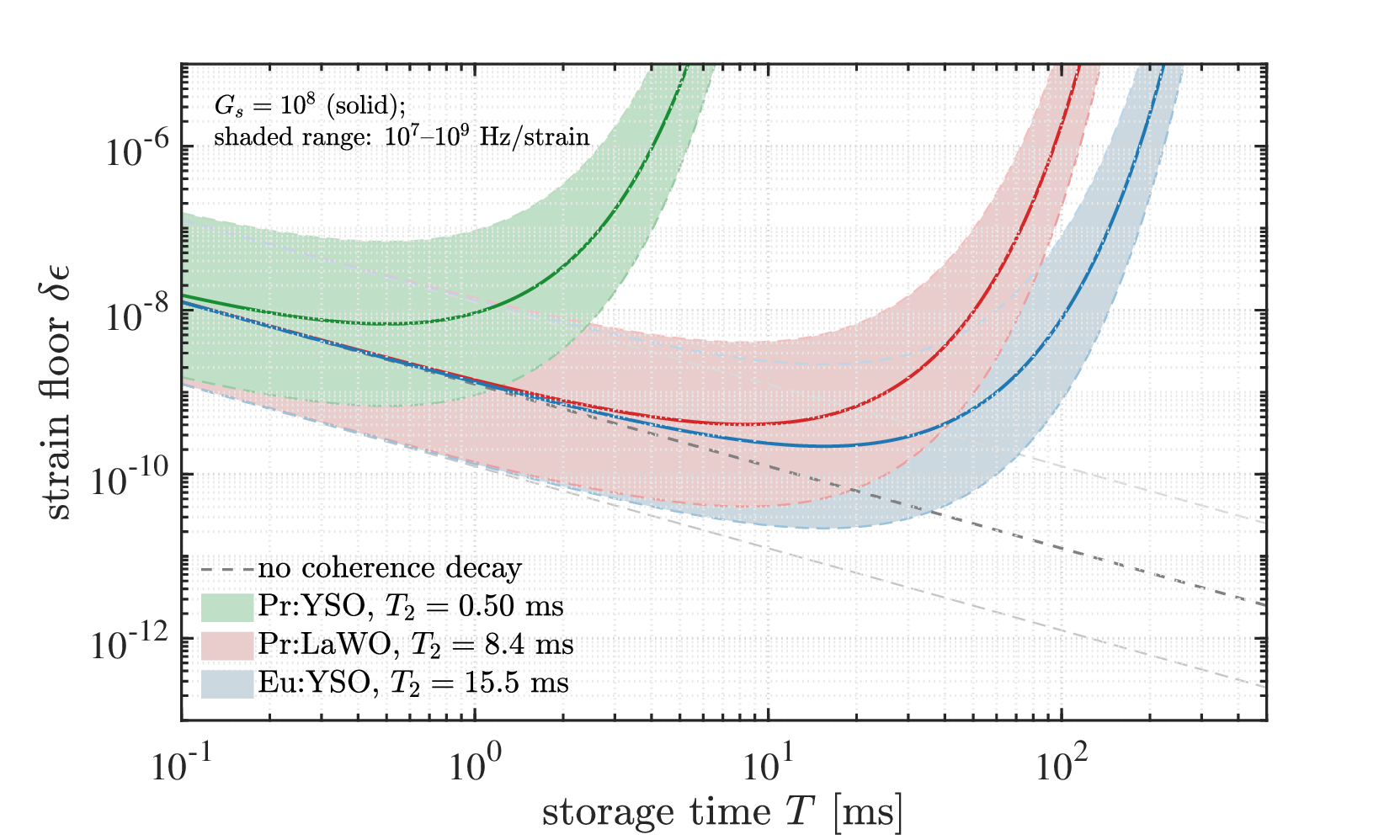}
\caption{\rev{Single-shot strain sensitivity in the presence of coherence decay. The grey dashed line shows the ideal limit without coherence decay. The coloured curves assume exponential coherence decay, $C(T)=\exp(-T/T_2)$, using representative zero-field coherence times for Pr:YSO ($T_2=0.50$ ms)~\cite{Heinze2014}, Pr:LaWO ($T_2=8.4$ ms)~\cite{Lovric2013}, and Eu:YSO ($T_2=15.5$ ms)~\cite{Alexander2007}. For Pr:LaWO, $T_2=8.4$ ms is the coherence-amplitude lifetime corresponding to the measured $4.2$ ms $1/e$ retrieval intensity decay under CPMG. Solid curves use $G_s=10^8$ Hz/strain, while the shaded regions span $G_s=10^7$--$10^9$ Hz/strain. We assume $N_{\rm sb}(0)=10^6$, $\mathcal V=1$, and ideal homodyne readout.}}
\label{fig:sensitivity}
\end{figure}

\rev{Figure~\ref{fig:sensitivity} shows the matched-DD optical shot-noise strain floor as a function of storage time. For fixed $G_s$ and $N_{\rm sb}(0)$, the sensitivity initially improves through phase accumulation and then worsens once the loss of coherence outweighs the gain in interrogation time. The grey dashed curve is the no-decay limit. Solid curves use $G_s=10^8$ Hz/strain and the shaded regions span $10^7$--$10^9$ Hz/strain for the three zero-field coherence benchmarks above. The common value $N_{\rm sb}(0)=10^6$ is an assumed detected-sideband resource, not a measured value for all three experiments. The bands vary the assumed $|G_s|$.} Figure~\ref{fig:sensitivity} uses exponential envelopes, $C(T)=\exp(-T/T_2)$, with experimentally reported zero-field coherence scales in three rare-earth systems. The natural Pr:YSO value $T_2\simeq0.50$ ms is the established baseline quoted in Ref.~\cite{Heinze2014}. In Pr:La$_2$(WO$_4$)$_3$, Lovri\'c \emph{et al.} measured a 4.2-ms $1/e$ decay under CPMG, corresponding to an effective $T_2$ lifetime of $T_2=8.4$ ms~\cite{Lovric2013}. For $^{151}$Eu$^{3+}$:Y$_2$SiO$_5$, a zero-field Raman echo measurement gave $T_2=15.5\pm2$ ms \cite{Alexander2007}. These values are used only as experimentally relevant decay constants; they were obtained with different control sequences and do not define a common DD pulse number or sensing frequency. In an actual ac-strain measurement, $C(T)$ must be characterized for the chosen pulse timing and sequence.

\revstop

In conventional Raman heterodyne detection, the optical read field propagates through the crystal, generates the Raman sideband, and the transmitted carrier subsequently acts as the local oscillator for that sideband. Its intensity therefore cannot be increased independently of the interaction with the ions. For the optional phase-sensitive homodyne measurement considered in Fig.~\ref{fig:sensitivity}, a separate local oscillator can instead be added after the crystal, which does not perturb the ensemble.

To characterize this regime, we use the resonant saturation parameter $s_r=|\Omega_r|^2/(\Gamma_1\Gamma_2)=I_r/I_{\mathrm{sat}}$, where $\Gamma_1$ is the optical population decay rate and $\Gamma_2$ the optical coherence decay rate. In a steady-state closed two-level model, the excited-state population is $\rho_{ee}=s_r/[2(1+s_r)]$, while the power-broadened optical half-width becomes $\gamma_{\mathrm{pb}}=\Gamma_2\sqrt{1+s_r}$. Increasing the read intensity therefore increases both the excited-state population and the optical linewidth. In the weak-read regime, the Raman sideband photon number grows approximately linearly with $s_r$ for fixed read duration. As $s_r$ approaches unity, power broadening, optical pumping, population redistribution, reabsorption, and read-induced spin evolution become increasingly important, so the small-signal relation in Eq.~\eqref{eq:Esb} is no longer sufficient \cite{Kindem2018,FernandezGonzalvo2019}. We therefore treat $N_{\rm sb}$ as a measured readout quantity rather than infer it from the read carrier power. 

\section{Strain calibration, systematics, and material platforms}
\revstart
\label{sec:calibration}

\subsection{Anisotropic elastic transfer and spatial averaging}

A phase measurement does not by itself determine a strain coefficient: it measures the product of the hyperfine strain slope and the strain actually sampled by the ions. Relating that strain to an externally applied stress therefore requires the elastic response of the crystal. We use $\sigma_{ij}=C_{ijkl}\eps_{kl}$ and $\eps_{ij}=S_{ijkl}\sigma_{kl}$, where $i,j,k,l\in\{x,y,z\}$ denote Cartesian components, $C_{ijkl}$ is the stiffness tensor, and $S=C^{-1}$ is the corresponding compliance tensor. In a monoclinic host such as YSO, these tensors must be transformed from the crystallographic frame to the sample and actuator frame. The conversion cannot in general be described by a single Young modulus, and the boundary conditions imposed by the mount, adhesive, electrodes, and free surfaces must also be included.

For the six independent strain components $a\in\{xx,yy,zz,yz,xz,xy\}$ defined above, we write the local response to an actuator voltage $V_d(\omega)$ as
\begin{equation}
\eps_a(\bm r,\omega)
=
\Gamma_{aV}(\bm r,\omega)V_d(\omega),
\end{equation}
where $\Gamma_{aV}$ is the position- and frequency-dependent voltage-to-strain transfer function. It may be obtained from an independent calibration or from an elastic model that has been validated experimentally.

The spin ensemble experiences a spatially weighted average of this strain,
\begin{equation}
\bar\eps_a
=
\frac{\int w(\bm r)\eps_a(\bm r)\dd^3r}
{\int w(\bm r)\dd^3r},
\label{eq:strainavg}
\end{equation}
where $w(\bm r)$ describes the spatial weighting set by the participating ion ensemble and the optical and rf fields.

If the strain varies across the ensemble, the accumulated phase varies with position as well. This reduces the observed contrast according to
\begin{equation}
C_{\eps}
=
\left|
\frac{\int w(\bm r)\ee^{-\ii\Phi_{\eps}(\bm r)}\dd^3r}
{\int w(\bm r)\dd^3r}
\right|.
\end{equation}
An inhomogeneous stress field can therefore contribute both a mean phase shift and additional dephasing, consistent with mechanically induced inhomogeneous optical responses observed in rare-earth ensembles \cite{Zhang2020}. Where a calibrated hydrostatic or uniaxial pressure coefficient of an optical reference line is available, it can provide an additional calibration of the mechanical transfer \cite{Galland2020,Zhang2020}.
\subsection{Electric field and thermal systematics}

A voltage-driven actuator can expose the crystal to strain, electric field, and heating simultaneously. The measured transition shift should therefore be fitted as
\begin{equation}
\delta\nu_s=\sum_aG_{s,a}^{(\eps)}\bar\eps_a+\bm k_s^{(E)}\cdot\bar{\bm E}+k_s^{(T)}\delta T+\cdots,
\label{eq:systematics}
\end{equation}
with independently characterized transfer functions from drive voltage. Voltage reversal alone is not decisive because both piezo strain and electric field can be odd in voltage. Useful null tests are: compare the complex response on and off a mechanical resonance; vary shielding or electrode-crystal distance while preserving mechanical coupling; use a remote or non-electrical mechanical drive; calibrate the Stark response in a geometry with negligible strain; and compare transitions or site classes with different strain and electric field tensors. Temperature and laser frequency references should be interleaved with the phase measurement. The Eu:YSO Raman heterodyne Stark experiment provides a direct precedent for independently calibrating the electric field term \cite{Macfarlane2014}.

\subsection{Site symmetry and magnetically inequivalent subsites}

Rare-earth ions in YSO occupy two crystallographic sites, each with local $C_1$ symmetry \cite{Longdell2002,Longdell2006,Macfarlane2014}. Because $C_1$ has no nontrivial local symmetry operation, no component of a linear strain response is forbidden at an individual site. A particular transition or strain pattern can nevertheless have a vanishing projected coefficient. At a generic magnetic field orientation, each crystallographic site can form two magnetically inequivalent subsites related by the crystal twofold rotation $R$. Their parameters obey
\begin{equation}
\bm B'=R\bm B,
\qquad
\eps'=R\eps R^T,
\end{equation}
\begin{equation}
Q'=RQR^T,
\qquad
M'=RMR^T.
\end{equation}
These relations constrain the pair of transition slopes and reduce the number of independent global-fit parameters. At zero field, a symmetry-preserving strain pattern may leave the two orientations degenerate, whereas a generic field or strain can resolve them. An experiment must therefore state whether crystallographic sites and magnetic subsites are spectrally resolved, separately fitted, or deliberately averaged.

\section{From scalar phase slopes to effective and microscopic tensors}
\label{sec:tensors}

The first quantity to determine experimentally would be the effective strain Hamiltonian of the hyperfine manifold. For each calibrated strain channel $a$, collect transition frequencies and slopes $G_{s,a}$ over several transitions, orientations, and, where useful, small magnetic fields. Numerically diagonalizing Eq.~\eqref{eq:Heff} for trial tensors gives the fit model
\begin{equation}
\nu_s^{\mathrm{mod}}(\bm B,\eps;\bm\vartheta)=\lambda_s(\bm B,\eps;\bm\vartheta)-\lambda_g(\bm B,\eps;\bm\vartheta),
\end{equation}
where $\lambda_g$ and $\lambda_s$ are eigenvalues of $H_s/h$ and $\bm\vartheta$ contains $Q^{(0)}$, $M^{(0)}$, and the effective derivatives $K^{(Q,a)}$ and $K^{(M,a)}$. The data vector may include both transition frequencies and slopes obtained from the phase measurements,
\begin{equation}
\bm d=\{\nu_s,\partial_a\nu_s\},
\qquad
\chi^2=(\bm d-\bm m)^T\Sigma^{-1}(\bm d-\bm m).
\end{equation}
The Jacobian $J_{\mu k}=\partial m_\mu/\partial\vartheta_k$ shows whether the measured observables contain independent information about the fitted parameters. If its columns are not linearly independent, some parameter combinations cannot be distinguished from the available data. At zero field, an isotropic part of $K^{(Q,a)}$ adds only a state-independent constant and is not spectroscopically visible; each strain channel therefore contributes at most five independent traceless symmetric quadrupole-like parameters. \rev{This parameter count is not a guarantee of full identifiability from zero-field frequency shifts: an $I=5/2$ manifold has only three hyperfine doublets and two independent transition frequencies. At first order, changes that only rotate the quadrupole principal axes are invisible to these frequencies. Recovering such components requires additional observables sensitive to the eigenstates or suitably oriented finite-field data; the latter also introduce the derivatives $K^{(M,a)}$.}

Only after the effective tensors are constrained should one fit the microscopic $\gamma_a^{kq}$ in Eq.~\eqref{eq:Hstrain}. That second stage inversion requires a crystal field model because $\partial_aQ$ and $\partial_aM$ depend on strain derivatives of both eigenvectors and energy denominators. Separating these two steps avoids interpreting a measured scalar $G_s$ as a single microscopic crystal field coefficient.

\section{Conclusion}
\revstart
In this work, we propose measuring the strain response of rare-earth ensembles through the phase accumulated by a long-lived hyperfine coherence, and retrieving this phase using Raman heterodyne detection. The system therefore acts as a coherent phase memory, converting strain-induced frequency shifts into an accumulated phase during storage and reading it out optically. This coherence is also used in spin-wave optical memories, ensuring compatibility with later operation using optical or nonclassical inputs.

Characterising the strain response of hyperfine levels in rare-earth systems is challenging. The response is generally anisotropic, with several strain components often present simultaneously, while synchronous electric field or thermal shifts can mimic the desired signal. Rather than relying solely on instantaneous spectral displacement, we propose measuring the strain-induced phase over a controlled sensing interval. Combined with transition selection and the filter function of the DD sequence, this provides a means of constraining strain coefficients that are otherwise difficult to access directly. Raman heterodyne readout has already been demonstrated to detect sub-Hertz changes in rare-earth hyperfine transitions. Extending this to strain requires calibration of the transfer function from actuator voltage to strain and separation of strain from synchronous parasitic shifts. Measurements over different transitions and strain geometries can then constrain the accessible effective strain coefficients. 

The proposed method therefore provides a practical route to characterising strain couplings through coherent phase measurements. \textcolor{black}{Spin-based detection of strain and mechanical motion has close precedents in diamond defect systems \cite{Ovartchaiyapong2014,Teissier2014}; here the sensing degree of freedom is instead a rare-earth hyperfine coherence.} The same phase-sensitive approach can be extended to narrowband strain spectroscopy, coherent acoustic-mode detection, and rare-earth optomechanics \cite{Molmer2016,Ohta2021,Louchet2023}.

\revstop

\begin{acknowledgments}
I thank Ceyhun Bulutay for valuable comments on an earlier version of the manuscript and the anonymous referee(s) for constructive suggestions, in particular for pointing out the practical advantages of Raman heterodyne readout over an optical input--output memory implementation and for noting the useful comparison of the scaling estimate in Eq.~\eqref{eq:stark_scaling} with the experimental results of Ref.~\cite{Macfarlane2014}. This work is supported by the Einstein Foundation Berlin through an Independent Researcher Grant, and by DLR through funds provided by BMFTR (OPTIMUS PRIME, No. 50SI2655A). ChatGPT (OpenAI), including GPT-5.2 Pro and GPT-6 Astra Pro, was used as a supporting tool for analytical calculations, simulations and code development, and manuscript editing. The scientific conception, analysis, interpretation, and conclusions were developed by the author. Calculations were independently checked by the author, who takes full responsibility for the scientific content of the manuscript.
\end{acknowledgments}

\section*{Data Availability}

The data that support the findings of this article are available from the author upon reasonable request.

\appendix

\section{How strain enters the effective hyperfine Hamiltonian}
\label{app:projection}

\subsection{Physical picture}

For the Pr$^{3+}$ and Eu$^{3+}$ systems considered here, the relevant electronic crystal field level is an isolated non-Kramers singlet. There is therefore no low-energy electronic doublet for strain to split directly. In addition, for an isolated time-reversal-symmetric singlet,
\begin{equation}
\bra{0}\bm J\ket{0}=0,
\end{equation}
so the magnetic hyperfine interaction and the electronic Zeeman interaction have no first-order matrix element within that singlet.

The nuclear-spin levels nevertheless depend on the electronic structure through both the local electric field gradient and virtual admixture of higher electronic states. The following compact derivation uses a fixed-$J$ multiplet. Two magnetic-hyperfine interactions generate the familiar pseudoquadrupole term, while one hyperfine and one electronic-Zeeman interaction generate the enhanced nuclear Zeeman response. Strain changes the energies and wave functions of the crystal field states and therefore changes these virtual processes. We collect these virtual processes in the tensor $\Lambda$.

It is useful to include strain in the electronic problem from the beginning. At fixed strain $\bm\eps$, define
\begin{equation}
H_{\mathrm{el}}(\bm\eps)
=H_{\mathrm{FI}}+H_{\mathrm{CF}}+H_{\eps},
\end{equation}
with
\begin{equation}
H_{\mathrm{el}}(\bm\eps)\ket{n(\bm\eps)}
=E_n(\bm\eps)\ket{n(\bm\eps)}.
\end{equation}
Let $P_0=\ket{0}\bra{0}$ project onto the selected electronic singlet and $Q_0=1-P_0$. We separate the remaining terms as
\begin{align}
U&=-g_n\mu_N\bm B\cdot\bm I+H_Q^{(n)},\\
W&=A_J\bm I\cdot\bm J+g_J\mu_B\bm B\cdot\bm J.
\end{align}
The terms in $U$ act directly on the nuclear spin. The terms in $W$ connect the selected singlet to other electronic crystal field states. Since $P_0WP_0=0$, their leading effect is second order. A L\"owdin or Schrieffer--Wolff projection at fixed strain gives \cite{Lowdin1951,SchriefferWolff1966}
\begin{equation}
\begin{aligned}
H_{\mathrm{eff}}={}&E_0P_0+P_0UP_0\\
&-P_0WQ_0
\frac{1}{Q_0H_{\mathrm{el}}Q_0-E_0}
Q_0WP_0+\cdots .
\end{aligned}
\end{equation}

\subsection{The tensors \texorpdfstring{$\Lambda$, $Q$, and $M$}{Lambda, Q, and M}}

Expanding the second-order term gives the two contributions relevant for the hyperfine transition. Two hyperfine interactions produce the pseudoquadrupole term, while the hyperfine--electronic-Zeeman cross term produces the enhanced nuclear Zeeman term. The electronic-Zeeman term squared gives a quadratic Zeeman shift that is independent of the nuclear spin and therefore does not change a transition within the chosen hyperfine manifold.

In frequency units, the effective Hamiltonian relevant to the transition is given by Eq.~\eqref{eq:Heff}, with
\begin{align}
M_{ij}(\bm\eps)
&=-\frac{g_n\mu_N}{h}\delta_{ij}
-\frac{2A_Jg_J\mu_B}{h}\Lambda_{ij}(\bm\eps),\\
Q_{ij}(\bm\eps)
&=Q_{ij}^{\mathrm{bare}}(\bm\eps)
-\frac{A_J^2}{h}\Lambda_{ij}(\bm\eps),
\end{align}
and
\begin{equation}
\Lambda_{ij}(\bm\eps)
=\sum_{n\neq0}
\frac{\bra{0}J_i\ket{n}\bra{n}J_j\ket{0}}
{E_n-E_0}.
\label{eq:Lambda_app}
\end{equation}
\rev{These are the standard fixed-$J$ effective-spin expressions \cite{Teplov1968,Longdell2002}. For Eu$^{3+}$ in its $^7F_0$ ground state, virtual magnetic-hyperfine coupling involves excited multiplets such as $^7F_1$: one must use the full electronic magnetic-hyperfine and magnetic-moment operators, rather than $A_J\bm J$ and $g_J\mu_B\bm J$ restricted to $J=0$ \cite{Longdell2006,Macfarlane2014}. The effective tensors $Q$ and $M$ and their strain derivatives still define the measured response; no fixed-$J$ numerical prediction for the Eu strain coefficient is made here. Equation~\eqref{eq:Lambda_app} can be read directly in terms of virtual transitions: each excited crystal field state provides one virtual path. The numerator says how strongly that state is connected to the selected singlet, while the denominator gives the energy cost of the virtual excitation.}

Strain changes both pieces of this sum. It shifts the crystal field energy spacings and it changes the crystal field wave functions. Neglecting an explicit strain dependence of $A_J$ and $g_J$ for the moment,
\begin{align}
\partial_a M_{ij}
&=-\frac{2A_Jg_J\mu_B}{h}\,\partial_a\Lambda_{ij},\\
\partial_a Q_{ij}
&=\partial_a Q_{ij}^{\mathrm{bare}}
-\frac{A_J^2}{h}\,\partial_a\Lambda_{ij}.
\label{eq:dQ_app}
\end{align}
Any known direct strain dependence of the hyperfine constant or other effective parameters can simply be added to these derivatives.

To separate the two contributions to $\partial_a\Lambda_{ij}$, define
\begin{equation}
A_{ij}^{(n)}
=\bra{0}J_i\ket{n}\bra{n}J_j\ket{0},
\qquad
\Delta_n=E_n-E_0.
\end{equation}
Then
\begin{equation}
\partial_a\Lambda_{ij}
=\sum_{n\neq0}
\left[
\frac{\partial_a A_{ij}^{(n)}}{\Delta_n}
-\frac{A_{ij}^{(n)}}{\Delta_n^2}\,\partial_a\Delta_n
\right].
\label{eq:dLambda_app}
\end{equation}
The second term describes the strain-induced change of the energy denominator. By the Hellmann–Feynman theorem,
\begin{equation}
\partial_a\Delta_n
=\bra{n}V_a\ket{n}-\bra{0}V_a\ket{0}.
\label{eq:dDelta_app}
\end{equation}
The first term comes from strain-induced changes of the wave functions and therefore of the angular momentum matrix elements. For a nondegenerate crystal-field level,
\begin{equation}
\ket{\partial_a n}
=\sum_{m\neq n}\ket{m}
\frac{\bra{m}V_a\ket{n}}{E_n-E_m},
\label{eq:dn_app}
\end{equation}
and $\partial_a A_{ij}^{(n)}$ follows by differentiating its bra and ket factors. Thus strain modifies the virtual admixture through changes in both the level spacings and the matrix elements.

\subsection{Connection to the measured hyperfine shift}

At zero magnetic field the effective Hamiltonian reduces to
\begin{equation}
\frac{H_s}{h}=I_iQ_{ij}I_j.
\end{equation}
A small strain component $\eps_a$ therefore changes the hyperfine Hamiltonian by
\begin{equation}
\delta\!\left(\frac{H_s}{h}\right)
=\eps_a I_i(\partial_aQ_{ij})I_j.
\end{equation}
Taking the difference of this expectation value between the two sensing states gives the transition slope $G_{s,a}$ in Eq.~\eqref{eq:Gsa}. The virtual-admixture contribution follows the chain
\begin{equation}
\bm\eps
\;\longrightarrow\;
H_{\mathrm{CF}}
\;\longrightarrow\;
\Lambda
\;\longrightarrow\;
Q,M
\;\longrightarrow\;
\nu_s.
\end{equation}

A direct strain derivative of $Q^{\mathrm{bare}}$ provides an additional path to $\nu_s$. For the proposed experiment, the first useful quantity is the measured transition slope $G_s$, not an individual microscopic coefficient $\gamma_a^{kq}$. Recovering the microscopic crystal field strain coefficients requires the crystal field spectrum, matrix elements, and several independent transition and strain measurements, as discussed in Sec.~\ref{sec:tensors}.

This projection is specific to an isolated non-Kramers singlet. For a Kramers ion, or any system with a low-energy electronic doublet, strain can act directly inside that electronic manifold and the low-energy projection must be redone. The later phase accumulation and readout formalism is unchanged once the corresponding transition strain coefficient is known.

\section{RF write-pulse phase}
\revstart
\label{app:writephase}

Throughout this appendix, $p_g+p_s=1$ refers to the normalized population within the addressed two-level subensemble. Population outside the selected pair is not represented in the two-level density matrix and instead reduces the effective participating ensemble size $N_{\mathrm{eff}}$. With $P=p_g-p_s$ and $\sigma_z=\ket{g}\bra{g}-\ket{s}\bra{s}$, the initial Bloch vector is $\bm r_0=P\hat{\bm z}$. A rotation by angle $\theta$ about $\bm n=(\cos\phi_w,-\sin\phi_w,0)$ gives
\begin{equation}
\bm r=P[\hat{\bm z}\cos\theta+(\bm n\times\hat{\bm z})\sin\theta].
\end{equation}
Thus $r_x=-P\sin\phi_w\sin\theta$ and $r_y=-P\cos\phi_w\sin\theta$. Since $\rho_{gs}=(r_x-\ii r_y)/2$, Eq.~\eqref{eq:rhogs} follows. For positive $P$ and a $\pi/2$ write pulse, $\arg\rho_{gs}=\phi_w+\pi/2$. The sensing Hamiltonian in Sec.~\ref{sec:filter} then adds $+\Phi_{\eps}$ to this phase. A fixed quadrature offset is included in $\phi_0$; reversing a phase or state convention requires reversing the corresponding calibrated sign consistently.
\revstop

\section{Instantaneous-pulse filter function}

For $n_\pi$ ideal instantaneous $\pi$ pulses at times $0<t_1<\cdots<t_{n_\pi}<T$, define $t_0=0$ and $t_{n_\pi+1}=T$. The toggling function is constant between pulses and changes sign after each pulse, giving
\begin{equation}
Y(\omega)=\sum_{j=0}^{n_\pi}(-1)^j
\frac{\ee^{-\ii\omega t_{j+1}}-\ee^{-\ii\omega t_j}}
{-\ii\omega}.
\end{equation}
For CPMG timing $t_j=(j-1/2)T/n_\pi$. The expression above is the ideal-pulse result used to understand the passband analytically. For a pulse of finite duration, the spin is rotating while the pulse is applied and the exact sensitivity is given by the appropriate control frame modulation function rather than by a strict $y(t)=\pm1$ square wave \cite{Ishikawa2018}. The finite-width calculation underlying Fig.~\ref{fig:phase_accumulation} uses the common short-pulse approximation in which the sensing weight is set to zero during the rf pulse. This is a convenient approximation for the plotted timing comparison, not a universal property of finite pulses. The phase response to a deterministic strain tone depends on $Y(\omega)$ itself, whereas Gaussian frequency-noise dephasing depends on a spectral integral of $|Y(\omega)|^2$.

\section{Raman heterodyne and homodyne phase shot noise}
\revstart
The proposed optical readout is naturally a carrier--sideband heterodyne measurement. The Raman sideband beats with the transmitted read field at the hyperfine frequency, so the relative optical phase rotates during the detection window. In the strong-carrier limit, averaging the squared phase response over many beat cycles gives the phase Fisher information
\begin{equation}
F_\phi\simeq2\mathcal V^2\Nsb,
\end{equation}
and therefore
\begin{equation}
\delta\phi_{\mathrm{RH}}
=\frac{1}{\mathcal V\sqrt{2\Nsb}}.
\end{equation}
This is the conventional Raman heterodyne limit used when the carrier itself acts as the local oscillator. Selecting one electronic demodulation quadrature does not remove the usual heterodyne penalty.

For comparison, one may supply a separate phase sensitive local oscillator at the sideband frequency. The relative optical phase is then stationary. Near the quadrature point,
\begin{equation}
\left|\frac{\partial N}{\partial\phi}\right|
=2\mathcal V\sqrt{N_{\mathrm{LO}}\Nsb}.
\end{equation}
For $N_{\mathrm{LO}}\gg\Nsb$, the photon shot noise is $\sqrt{N_{\mathrm{LO}}}$, which gives the ideal homodyne limit
\begin{equation}
\delta\phi_{\mathrm{hom}}
=\frac{1}{2\mathcal V\sqrt{\Nsb}}.
\end{equation}
\rev{The conditional curves in Fig.~\ref{fig:sensitivity} use this ideal homodyne value;} conventional Raman heterodyne increases the optical noise-limited uncertainties by $\sqrt{2}$. Both expressions assume a coherent sideband and shot-noise-limited detection. Technical amplitude and phase noise must be measured independently.
\revstop

\section{RF and optical preparation of the sensing coherence}
\label{app:write}

The Raman sideband is proportional to the mean ground state coherence. Direct rf preparation can rotate the full optically prepared population difference into the transverse plane, whereas an optically written coherent spin wave generally corresponds to a much smaller Bloch sphere tilt~\cite{Sevincli2026}, which can then be read out relatively efficiently. \textcolor{black}{The two schemes therefore place the difficulty at different stages. A weak-excitation optical memory creates only a small collective spin tilt, but the memory control sequence can subsequently convert the stored excitation back into the desired optical mode with high efficiency. Direct rf preparation can instead create an equatorial spin coherence, but Raman heterodyne does not retrieve that coherence one-for-one as photons: the detected sideband has to be generated by the optical read field, and its strength is limited by the Raman conversion and by the onset of read-induced nonlinearities.}

Before the write pulse, let
\begin{equation}
\rho_0=\frac{1}{2}(\mathbf{1}+P\sigma_z),
\end{equation}
where $P$ is the population difference of the addressed hyperfine class. A resonant rf pulse of area $\theta$ and phase $\phi_w$ gives
\begin{equation}
\rho_{gs}(0^+)=
\frac{\ii P}{2}\ee^{\ii\phi_w}\sin\theta ,
\label{eq:rhogs}
\end{equation}
for the phase convention used in the main text. A $\pi/2$ pulse therefore produces
$|\rho_{gs}^{(\rm rf)}|=\frac{|P|}{2}$. Experimentally, scanning $\phi_w$ and verifying unit-slope transfer to the unwrapped Raman heterodyne phase provides a direct check of the preparation and readout chain.

For comparison, a phase-referenced coherent optical pulse mapped onto the same collective mode can be represented by a small tilt $\theta_{\rm spin}$ of the collective Bloch vector \cite{Sevincli2026}. In the weak excitation limit,
\begin{equation}
|\rho_{gs}^{(\rm opt)}|
\simeq
\frac{|P|}{2}\theta_{\rm spin},
\end{equation}
and the number of stored excitations is
\begin{equation}
\mu_{\rm st}
=
N_{\rm eff}\sin^2\!\left(\frac{\theta_{\rm spin}}{2}\right)
\simeq
\frac{N_{\rm eff}\theta_{\rm spin}^2}{4},
\label{eq:stored_tilt}
\end{equation}
with $N_{\rm eff}$ being the number of ions participating in the storage process. Thus even a small fractional excitation can correspond to a large absolute optical signal in an ensemble. For example, with $N_{\rm eff}=10^{10}$, tilts $\theta_{\rm spin}=0.01$, $0.03$, and $0.10$ rad correspond to approximately $2.5\times10^5$, $2.3\times10^6$, and $2.5\times10^7$ stored excitations, respectively. The corresponding minority-state populations are only $2.5\times10^{-5}$, $2.3\times10^{-4}$, and $2.5\times10^{-3}$. These numbers are illustrative; whether a particular optical memory implementation remains in its linear EIT or Raman regime must still be checked from its optical depth and control field parameters.

If an effective fraction $\eta_{\rm out}$ of the stored excitation is retrieved and detected after storage, the detected photon number is approximately $N_{\rm det}\simeq \eta_{\rm out}\mu_{\rm st}$.
Taking $\eta_{\rm out}=0.03$ as an illustrative post-storage retrieval and detection efficiency gives about $7.5\times10^3$, $6.8\times10^4$, and $7.5\times10^5$ detected photons for the three tilts above. \textcolor{black}{The value $0.03$ is only an illustrative detected-output fraction for this estimate; it is not meant as a typical upper limit for optical-memory retrieval, which can be much larger in optimized rare-earth memories \cite{Hedges2010}.} For conventional Raman heterodyne, the corresponding photon-shot-noise phase floors,
\begin{equation}
\delta\phi_{\rm RH}\simeq\frac{1}{\sqrt{2N_{\rm det}}},
\end{equation}
are approximately $8.2$, $2.7$, and $0.82$ mrad. Thus a small optical tilt can already provide a measurable phase signal even when only a few percent of the stored excitation is detected.

\textcolor{black}{Optical storage produces a relatively small spin excitation that can nevertheless be retrieved efficiently, whereas rf preparation creates a much larger spin coherence but relies on a weaker, driven Raman conversion for readout.} In the perturbative Raman heterodyne regime,
$N_{\rm sb}\propto s_r\,|\rho_{gs}|^2$, where
$s_r=\frac{|\Omega_r|^2}{\Gamma_1\Gamma_2}
\simeq\frac{I_r}{I_{\rm sat}}$
is the read field saturation parameter. Combining this scaling with the coherence amplitudes above gives
\begin{equation}
\frac{N_{\rm sb}^{(\rm rf)}}{N_{\rm sb}^{(\rm opt)}}
\simeq
\frac{1}{\theta_{\rm spin}^2}
\end{equation}
at the same read intensity, or equivalently
\begin{equation}
s_r^{(\rm rf)}
\simeq
\theta_{\rm spin}^2 s_r^{(\rm opt)}
\label{eq:read_reduction}
\end{equation}
for the same sideband photon number.

For $\theta_{\rm spin}=0.03$, Eq.~\eqref{eq:read_reduction} gives a reduction of the required read saturation parameter by roughly $10^3$. A readout that would require $s_r\sim1$ for the small optical coherence would therefore require only $s_r\sim9\times10^{-4}$ after rf $\pi/2$ preparation. At such values the excited state population and power broadening are both negligible in the simple two-level estimate. Beyond the experimental simplicity with respect to an optical memory experiment, another advantage of RF preparation is that a given sideband signal can be obtained at much lower read intensity, keeping the readout in the perturbative regime. 

\bibliography{references}

@incollection{Goldner2015,
  author = {P. Goldner and A. Ferrier and O. Guillot-No\"el},
  title = {Rare Earth-Doped Crystals for Quantum Information Processing},
  booktitle = {Handbook on the Physics and Chemistry of Rare Earths},
  volume = {46},
  editor = {J.-C.~G.~B\"unzli and V.~K.~Pecharsky},
  publisher = {Elsevier},
  address = {Amsterdam},
  pages = {1-78},
  year = {2015},
  doi = {10.1016/B978-0-444-63260-9.00267-4}
}

@article{Moldes2026,
  title = {Long-Lived Telecom-Heralded Single-Photon Storage in an Absorptive Spin-Rephased Quantum Memory},
  author = {Rodr\'{\i}guez-Moldes, Alberto E. and Appas, F\'elicien and H\"anni, Jonathan and Rakonjac, Jelena V. and Grandi, Samuele and de Riedmatten, Hugues},
  journal = {Phys. Rev. Lett.},
  volume = {137},
  issue = {12},
  pages = {120803},
  numpages = {8},
  year = {2026},
  month = {Sep},
  publisher = {American Physical Society},
  doi = {10.1103/ftkb-pkvp},
  url = {https://link.aps.org/doi/10.1103/ftkb-pkvp}
}

@article{Gundogan2026_velo,
    author = {Gündoğan, Mustafa and Ahmadi, Arash and Krutzik, Markus},
    title = {Memory-assisted squeezed light velocimetry under realistic loss and incoherent noise},
    journal = {AVS Quantum Science},
    volume = {8},
    number = {3},
    pages = {034405},
    year = {2026},
    month = {09},
    issn = {2639-0213},
    doi = {10.1116/5.0345557},
    url = {https://doi.org/10.1116/5.0345557},
}

@misc{Videos_SM,
  note = {See Supplemental Material at [URL will be inserted by publisher]
  for Videos 1 and 2 showing the individual pseudospins and collective
  Bloch vector during the strain-induced phase evolution for the
  synchronized and 5\% detuned dynamical-decoupling sequences.}
}

@article{Heinze2014,
  author = {G. Heinze and C. Hubrich and T. Halfmann},
  title = {Coherence time extension in Pr$^{3+}$:Y$_2$SiO$_5$ by self-optimized magnetic fields and dynamical decoupling},
  journal = {Phys. Rev. A},
  volume = {89},
  pages = {053825},
  year = {2014},
  doi = {10.1103/PhysRevA.89.053825}
}

@article{Anai2024,
  author = {K. Anai and Y. Enomoto and H. Omura and K. Nagano and K. Izumi and M. Endo and S. Takeda},
  title = {Quantum-enhanced optical phase-insensitive heterodyne detection beyond 3-dB noise penalty of image band},
  journal = {Opt. Express},
  volume = {32},
  number = {11},
  pages = {19372--19387},
  year = {2024},
  doi = {10.1364/OE.498691}
}

@article{Alexander2007,
  author = {A. L. Alexander and J. J. Longdell and M. J. Sellars},
  title = {Measurement of the ground-state hyperfine coherence time of $^{151}$Eu$^{3+}$:Y$_2$SiO$_5$},
  journal = {J. Opt. Soc. Am. B},
  volume = {24},
  pages = {2479--2482},
  year = {2007},
  doi = {10.1364/JOSAB.24.002479}
}

@article{Hedges2010,
  author = {M. P. Hedges and J. J. Longdell and Y. Li and M. J. Sellars},
  title = {Efficient quantum memory for light},
  journal = {Nature},
  volume = {465},
  pages = {1052},
  year = {2010},
  doi = {10.1038/nature09081}
}

@article{Gundogan2015,
  author = {M. G\"undo\u{g}an and P. M. Ledingham and K. Kutluer and M. Mazzera and H. de Riedmatten},
  title = {Solid State Spin-Wave Quantum Memory for Time-Bin Qubits},
  journal = {Phys. Rev. Lett},
  volume = {114},
  pages = {230501},
  year = {2015},
  doi = {10.1103/PhysRevLett.114.230501}
}

@article{Lovric2013,
  author = {M. Lovri\'c and D. Suter and A. Ferrier and P. Goldner},
  title = {Faithful Solid State Optical Memory with Dynamically Decoupled Spin Wave Storage},
  journal = {Phys. Rev. Lett},
  volume = {111},
  pages = {020503},
  year = {2013},
  doi = {10.1103/PhysRevLett.111.020503}
}

@article{Zhong2015,
  author = {M. Zhong and M. P. Hedges and R. L. Ahlefeldt and J. G. Bartholomew and S. E. Beavan and S. M. Wittig and J. J. Longdell and M. J. Sellars},
  title = {Optically addressable nuclear spins in a solid with a six-hour coherence time},
  journal = {Nature},
  volume = {517},
  pages = {177},
  year = {2015},
  doi = {10.1038/nature14025}
}

@article{Ma2021,
  author = {Y. Ma and Y.-Z. Ma and Z.-Q. Zhou and C.-F. Li and G.-C. Guo},
  title = {One-hour coherent optical storage in an atomic frequency comb memory},
  journal = {Nat. Commun},
  volume = {12},
  pages = {2381},
  year = {2021},
  doi = {10.1038/s41467-021-22706-y}
}

@Article{Ortu2022,
author={Ortu, Antonio
and Holz{\"a}pfel, Adrian
and Etesse, Jean
and Afzelius, Mikael},
title={Storage of photonic time-bin qubits for up to 20{\thinspace}ms in a rare-earth doped crystal},
journal={npj Quantum Information},
year={2022},
month={Mar},
day={15},
volume={8},
number={1},
pages={29},
issn={2056-6387},
doi={10.1038/s41534-022-00541-3},
}

@article{Macfarlane2002,
  author = {R. M. Macfarlane},
  title = {High-resolution laser spectroscopy of rare-earth doped insulators: a personal perspective},
  journal = {J. Lumin},
  volume = {100},
  pages = {1},
  year = {2002},
  doi = {10.1016/S0022-2313(02)00450-7}
}

@article{Galland2020,
  title = {Mechanical Tunability of an Ultranarrow Spectral Feature of a Rare-Earth-Doped Crystal via Uniaxial Stress},
  author = {Galland, N. and Lu\ifmmode \check{c}\else \v{c}\fi{}i\ifmmode \acute{c}\else \'{c}\fi{}, N. and Fang, B. and Zhang, S. and Le Targat, R. and Ferrier, A. and Goldner, P. and Seidelin, S. and Le Coq, Y.},
  journal = {Phys. Rev. Appl.},
  volume = {13},
  issue = {4},
  pages = {044022},
  numpages = {7},
  year = {2020},
  month = {Apr},
  publisher = {American Physical Society},
  doi = {10.1103/PhysRevApplied.13.044022},
}

@article{Zhang2020,
  author = {S. Zhang and N. Galland and N. Lu\v{c}i\'c and R. Le Targat and A. Ferrier and P. Goldner and B. Fang and Y. Le Coq and S. Seidelin},
  title = {Inhomogeneous response of an ion ensemble from mechanical stress},
  journal = {Phys. Rev. Research},
  volume = {2},
  pages = {013306},
  year = {2020},
  doi = {10.1103/PhysRevResearch.2.013306}
}

@article{Molmer2016,
  title = {Dispersive coupling between light and a rare-earth-ion--doped mechanical resonator},
  author = {M\o{}lmer, Klaus and Le Coq, Yann and Seidelin, Signe},
  journal = {Phys. Rev. A},
  volume = {94},
  issue = {5},
  pages = {053804},
  numpages = {6},
  year = {2016},
  month = {Nov},
  publisher = {American Physical Society},
  doi = {10.1103/PhysRevA.94.053804},
}

@article{Ohta2021,
  author = {R. Ohta and L. Herpin and E. M. Weig and H. Yamaguchi and H. Okamoto},
  title = {Rare-earth-mediated opto-mechanical system in the reversed dissipation regime},
  journal = {Phys. Rev. Lett},
  volume = {126},
  pages = {047404},
  year = {2021},
  doi = {10.1103/PhysRevLett.126.047404}
}

@article{Louchet2023,
  author = {A. Louchet-Chauvet and P. Verlot and J.-P. Poizat and T. Chaneli\`ere},
  title = {Piezo-orbital backaction force in a rare-earth-doped crystal},
  journal = {Phys. Rev. Applied},
  volume = {20},
  pages = {054004},
  year = {2023},
  doi = {10.1103/PhysRevApplied.20.054004}
}

@article{deLange2011,
  author = {G. de Lange and D. Rist\`e and V. V. Dobrovitski and R. Hanson},
  title = {Single-spin magnetometry with multipulse sensing sequences},
  journal = {Phys. Rev. Lett},
  volume = {106},
  pages = {080802},
  year = {2011},
  doi = {10.1103/PhysRevLett.106.080802}
}

@article{Ovartchaiyapong2014,
  author = {P. Ovartchaiyapong and K. W. Lee and B. A. Myers and A. C. Bleszynski Jayich},
  title = {Dynamic strain-mediated coupling of a single diamond spin to a mechanical resonator},
  journal = {Nat. Commun},
  volume = {5},
  pages = {4429},
  year = {2014},
  doi = {10.1038/ncomms5429}
}

@article{Teissier2014,
  author = {J. Teissier and A. Barfuss and P. Appel and E. Neu and P. Maletinsky},
  title = {Strain coupling of a nitrogen-vacancy center spin to a diamond mechanical oscillator},
  journal = {Phys. Rev. Lett},
  volume = {113},
  pages = {020503},
  year = {2014},
  doi = {10.1103/PhysRevLett.113.020503}
}

@article{Degen2017,
  author = {C. L. Degen and F. Reinhard and P. Cappellaro},
  title = {Quantum sensing},
  journal = {Rev. Mod. Phys},
  volume = {89},
  pages = {035002},
  year = {2017},
  doi = {10.1103/RevModPhys.89.035002}
}

@article{Cywinski2008,
  author = {{\L}ukasz Cywi\'nski and R. M. Lutchyn and C. P. Nave and S. Das Sarma},
  title = {How to enhance dephasing time in superconducting qubits},
  journal = {Phys. Rev. B},
  volume = {77},
  pages = {174509},
  year = {2008},
  doi = {10.1103/PhysRevB.77.174509}
}

@article{Ishikawa2018,
  author = {T. Ishikawa and A. Yoshizawa and Y. Mawatari and S. Kashiwaya and H. Watanabe},
  title = {Influence of dynamical decoupling sequences with finite-width pulses on quantum sensing for AC magnetometry},
  journal = {Phys. Rev. Applied},
  volume = {10},
  pages = {054059},
  year = {2018},
  doi = {10.1103/PhysRevApplied.10.054059}
}

@article{Longdell2002,
  author = {J. J. Longdell and M. J. Sellars and N. B. Manson},
  title = {Hyperfine interaction in ground and excited states of praseodymium-doped yttrium orthosilicate},
  journal = {Phys. Rev. B},
  volume = {66},
  pages = {035101},
  year = {2002},
  doi = {10.1103/PhysRevB.66.035101}
}

@article{Longdell2006,
  author = {J. J. Longdell and A. L. Alexander and M. J. Sellars},
  title = {Characterization of the hyperfine interaction in europium-doped yttrium orthosilicate and europium chloride hexahydrate},
  journal = {Phys. Rev. B},
  volume = {74},
  pages = {195101},
  year = {2006},
  doi = {10.1103/PhysRevB.74.195101}
}

@article{GuillotNoel2009,
  author = {O. Guillot-No\"el and others},
  title = {Hyperfine structure and hyperfine coherent properties of praseodymium in single-crystalline La$_2$(WO$_4$)$_3$ by hole-burning and photon-echo techniques},
  journal = {Phys. Rev. B},
  volume = {79},
  pages = {155119},
  year = {2009},
  doi = {10.1103/PhysRevB.79.155119}
}

@article{Stevens1952,
  author = {K. W. H. Stevens},
  title = {Matrix elements and operator equivalents connected with the magnetic properties of rare earth ions},
  journal = {Proc. Phys. Soc. A},
  volume = {65},
  pages = {209},
  year = {1952},
  doi = {10.1088/0370-1298/65/3/308}
}

@incollection{Hutchings1964,
  author = {M. T. Hutchings},
  title = {Point-charge calculations of energy levels of magnetic ions in crystalline electric fields},
  booktitle = {Solid State Physics},
  volume = {16},
  editor = {F. Seitz and D. Turnbull},
  publisher = {Academic Press},
  address = {New York},
  pages = {227-273},
  year = {1964},
  doi = {10.1016/S0081-1947(08)60517-2}
}

@article{Lowdin1951,
  author = {P.-O. L\"owdin},
  title = {A note on the quantum-mechanical perturbation theory},
  journal = {J. Chem. Phys},
  volume = {19},
  pages = {1396},
  year = {1951},
  doi = {10.1063/1.1748067}
}

@article{SchriefferWolff1966,
  author = {J. R. Schrieffer and P. A. Wolff},
  title = {Relation between the Anderson and Kondo Hamiltonians},
  journal = {Phys. Rev},
  volume = {149},
  pages = {491},
  year = {1966},
  doi = {10.1103/PhysRev.149.491}
}

@article{Fraval2004,
  author = {E. Fraval and M. J. Sellars and J. J. Longdell},
  title = {Method of extending hyperfine coherence times in Pr$^{3+}$:Y$_2$SiO$_5$},
  journal = {Phys. Rev. Lett},
  volume = {92},
  pages = {077601},
  year = {2004},
  doi = {10.1103/PhysRevLett.92.077601}
}

@article{Heinze2013,
  author = {G. Heinze and C. Hubrich and T. Halfmann},
  title = {Stopped light and image storage by electromagnetically induced transparency up to the regime of one minute},
  journal = {Phys. Rev. Lett},
  volume = {111},
  pages = {033601},
  year = {2013},
  doi = {10.1103/PhysRevLett.111.033601}
}

@article{Souza2011,
  title = {Robust Dynamical Decoupling for Quantum Computing and Quantum Memory},
  author = {Souza, Alexandre M. and \'Alvarez, Gonzalo A. and Suter, Dieter},
  journal = {Phys. Rev. Lett.},
  volume = {106},
  issue = {24},
  pages = {240501},
  numpages = {4},
  year = {2011},
  month = {Jun},
  publisher = {American Physical Society},
  doi = {10.1103/PhysRevLett.106.240501},
}

@article{Macfarlane2014,
  author = {R. M. Macfarlane and A. Arcangeli and A. Ferrier and P. Goldner},
  title = {Optical Measurement of the Effect of Electric Fields on the Nuclear Spin Coherence of Rare-Earth Ions in Solids},
  journal = {Phys. Rev. Lett.},
  volume = {113},
  pages = {157603},
  year = {2014},
  doi = {10.1103/PhysRevLett.113.157603}
}

@article{Teplov1968,
  author = {M. A. Teplov},
  journal = {Sov. Phys. JETP},
  volume = {26},
  pages = {872},
  year = {1968}
}

@article{Mims1968,
  author = {W. B. Mims},
  title = {Phase Memory in Electron Spin Echoes, Lattice Relaxation Effects in {CaWO$_4$:Er,Ce,Mn}},
  journal = {Phys. Rev.},
  volume = {168},
  pages = {370--389},
  year = {1968},
  doi = {10.1103/PhysRev.168.370}
}

@article{Mossberg1982,
  author = {T. W. Mossberg},
  title = {Time-domain frequency-selective optical data storage},
  journal = {Opt. Lett.},
  volume = {7},
  pages = {77--79},
  year = {1982},
  doi = {10.1364/OL.7.000077}
}

@article{Mlynek1983,
  author = {J. Mlynek and N. C. Wong and R. G. DeVoe and E. S. Kintzer and R. G. Brewer},
  title = {Raman heterodyne detection of nuclear magnetic resonance},
  journal = {Phys. Rev. Lett.},
  volume = {50},
  pages = {993--996},
  year = {1983},
  doi = {10.1103/PhysRevLett.50.993}
}

@article{Wong1983,
  author = {N. C. Wong and E. S. Kintzer and J. Mlynek and R. G. DeVoe and R. G. Brewer},
  title = {Raman heterodyne detection of nuclear magnetic resonance},
  journal = {Phys. Rev. B},
  volume = {28},
  pages = {4993--5010},
  year = {1983},
  doi = {10.1103/PhysRevB.28.4993}
}

@article{Mitsunaga1985,
  author = {M. Mitsunaga and E. S. Kintzer and R. G. Brewer},
  title = {Raman heterodyne interference: Observations and analytic theory},
  journal = {Phys. Rev. B},
  volume = {31},
  pages = {6947--6960},
  year = {1985},
  doi = {10.1103/PhysRevB.31.6947}
}

@article{Holliday1990,
  author = {K. Holliday and X.-F. He and P. T. H. Fisk and N. B. Manson},
  title = {Raman heterodyne detection of electron paramagnetic resonance},
  journal = {Opt. Lett.},
  volume = {15},
  pages = {983--985},
  year = {1990},
  doi = {10.1364/OL.15.000983}
}

@article{Erickson1990,
  author = {L. E. Erickson},
  title = {Optical-pumping effects on Raman-heterodyne-detected multipulse rf nuclear-spin-echo decay},
  journal = {Phys. Rev. B},
  volume = {42},
  pages = {3789--3796},
  year = {1990},
  doi = {10.1103/PhysRevB.42.3789}
}

@article{FernandezGonzalvo2019,
  author = {X. Fernandez-Gonzalvo and S. P. Horvath and Y.-H. Chen and J. J. Longdell},
  title = {Cavity-enhanced Raman heterodyne spectroscopy in {Er$^{3+}$:Y$_2$SiO$_5$} for microwave-to-optical signal conversion},
  journal = {Phys. Rev. A},
  volume = {100},
  pages = {033807},
  year = {2019},
  doi = {10.1103/PhysRevA.100.033807}
}

@article{Kindem2018,
  author = {J. M. Kindem and J. G. Bartholomew and P. J. T. Woodburn and T. Zhong and I. Craiciu and R. L. Cone and C. W. Thiel and A. Faraon},
  title = {Characterization of $^{171}$Yb$^{3+}$:YVO$_4$ for photonic quantum technologies},
  journal = {Phys. Rev. B},
  volume = {98},
  pages = {024404},
  year = {2018},
  doi = {10.1103/PhysRevB.98.024404}
}

@article{King2021,
  author = {G. G. G. King and P. S. Barnett and J. G. Bartholomew and A. Faraon and J. J. Longdell},
  title = {Probing strong coupling between a microwave cavity and a spin ensemble with Raman heterodyne spectroscopy},
  journal = {Phys. Rev. B},
  volume = {103},
  pages = {214305},
  year = {2021},
  doi = {10.1103/PhysRevB.103.214305}
}

@article{Sevincli2026,
  title = {Generation of squeezed optical states via stored classical pulses in a Bose gas},
  author = {Sevin\ifmmode \mbox{\c{c}}\else \c{c}\fi{}li, Sevilay and R\"atzel, Dennis and Krutzik, Markus and Oktel, Mehmet \"Ozg\"ur and G\"undo\ifmmode \breve{g}\else \u{g}\fi{}an, Mustafa},
  journal = {Phys. Rev. Res.},
  volume = {8},
  issue = {3},
  pages = {033016},
  numpages = {13},
  year = {2026},
  month = {Jul},
  publisher = {American Physical Society},
  doi = {10.1103/89xy-86yz},
}

\end{document}